\documentclass[a4paper,11pt]{article}
\usepackage{pos}
\usepackage{graphicx}
\usepackage{subcaption}
\usepackage{wrapfig}
\usepackage{gensymb}

\newcommand{\U}[1]{\ensuremath{\mathrm{\ #1}}}
\newcommand{\UU}[2]{\ensuremath{\mathrm{\ #1^{#2}}}}

\title{A Search for High-Energy Emission from the Remnant of Supernova 1181 }
 \ShortTitle{SN 1181}

\author*[a]{J. Holder}
\author[]{for the VERITAS Collaboration}
\author[]{\hspace{-0.4em}}
\author[b]{Qiqi Jiang}
\affiliation[a]{Department of Physics and Astronomy and the Bartol Research Institute, \\ University of Delaware, Newark, DE 19716, USA}
\affiliation[b]{Institute of Physics and Astronomy, University of Potsdam, 14476 Potsdam-Golm, Germany and DESY, Platanenallee 6, 15738 Zeuthen, Germany}

\emailAdd{jholder@udel.edu}

\abstract{Over the previous millennium, only five Galactic supernovae were observed and recorded by contemporary astronomers, and their current-day counterparts subsequently identified. The remnants of four of these have all been very deeply studied, and ultimately detected, by TeV instruments after exposures of typically hundreds of hours. The measured TeV fluxes range from 1 Crab (by definition) down to 0.3\% Crab. The location of the fifth supernova remnant tied to a historical record of its supernova (SN 1181) has never been studied at TeV energies. The reason for this is simple – the associated remnant was only identified as such in 2021. The remnant, Pa\,30, is an unusual object whose properties are best explained as resulting from a Type Iax supernova explosion. These are a rare sub-type of Type Ia supernovae in which the merging white dwarfs are not fully destroyed by the supernova explosion, leading to a double-degenerate merger product colorfully described as a “zombie star”. We will present the results of a search for TeV gamma-ray emission from Pa\,30 with VERITAS.}

\ConferenceLogo{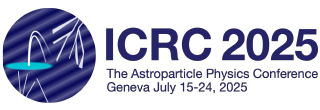}

\FullConference{39th International Cosmic Ray Conference (ICRC2025)\\
 15–24 July 2025\\
Geneva, Switzerland\\}

\begin{document}
\maketitle

\section{Introduction}
Supernova remnants have long been considered potential sites for cosmic ray acceleration and associated very high energy (VHE) gamma-ray emission (e.g. \citep{1994A&A...287..959D}). The remnants of historical Galactic supernovae are particularly good targets: contemporaneous observations of the supernovae mean that the age of the remnants is very accurately known, and give clues to the nature of their progenitors. In addition, they are extremely well-studied at all wavelengths. The locations of historical supernovae have therefore been observed with exposures of hundreds or thousands of hours with ground-based gamma-ray observatories and, in most cases, subsequently detected. These include SN 1006 (1006 AD) \citep{2010A&A...516A..62A}, Tycho's SNR (1572 AD) \citep{2011ApJ...730L..20A}, Kepler's SNR (1604 AD) \citep{2022A&A...662A..65H} and the Crab (1054 AD) \citep{1989ApJ...342..379W}. In the case of the Crab the bright VHE gamma-ray emission is due to the pulsar wind nebula powered by the Crab pulsar, while for the other objects the emission is much fainter, and is the result of leptonic or hadronic particle acceleration in the supernova remnant shocks. VHE emission has also been discovered from Cassiopeia A \citep{2001A&A...370..112A} the remnant of a supernova which likely occurred in the mid-1600's, although no definitive historical observations of this event exist.

Absent from this list is the remnant of SN\,1181, a historical supernova which was clearly described in contemporaneous Chinese and Japanese records. The supernova was visible as a "Guest Star" in the sky for 185 days, from 1181 August 6 to 1182 February 6, with no indication of motion on the celestial sphere. Modern-day searches initially identified 3C\,58 as a possible counterpart \citep{2002ISAA....5.....S}. 3C\,58 is a bright pulsar wind nebula powered by a pulsar (PSR\,J0205+6449) with extremely high spin-down power (5\% of the Crab). Gamma-ray observations focused on this object, eventually leading to a detection by MAGIC of a VHE gamma-ray source with a flux of just 0.65\% of the Crab Nebula \citep{2014A&A...567L...8A}. However, more recent studies suggest that the age of 3C\,58 lies in a range from $2400 - 7000\U{years}$ \citep{2008ApJS..174..379F}, significantly older than the expected age of $\sim850\U{years}$ for SN\,1181. 

\begin{figure}
  \begin{center}
    \begin{tabular}{c}
     \includegraphics[width=0.8\textwidth]{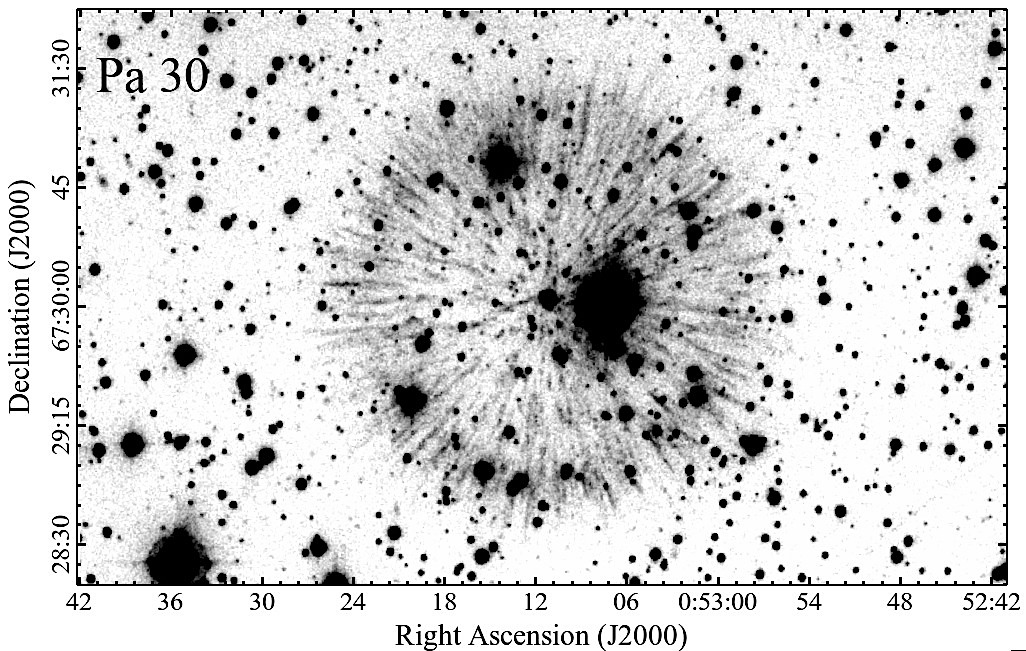}\\
    \end{tabular}
  \end{center}
  \caption  
      {
        \label{optical}
        Red [S II] image of Pa\,30. Figure from \citep{2023ApJ...945L...4F}. }
\end{figure}

Further re-evaluation now prefers an alternative counterpart to SN\,1181 known as Pa\,30 \citep{2021ApJ...918L..33R}. This object was first discovered by an amateur astronomer in 2013, in a search for planetary nebula candidates in WISE data \cite{2014apn6.confE..48K}. Figure~\ref{optical} shows a recent optical image of the nebula in the red [S II] band illustrating its extraordinary filamentary structure, with radial lines of emission converging on a hot central star \citep{2023ApJ...945L...4F}. The location of Pa\,30 is shown on the left of figure~\ref{maps}, along with revised constraints on the position of SN\,1181 from historical records \citep{2023MNRAS.523.3885S}. 3C\,58 is excluded as the counterpart by these constraints. 

At first glance, these results provide a compelling new target for high-energy follow-up observations. However, the remnant displays some unusual features which reduce the likelihood of it being a strong gamma-ray emitter. First among these is the fact that the central star cannot have resulted from a Type II (core collapse) or Type Ia (thermonuclear explosion of a CO white dwarf) supernova. Rather, it is believed to be the product of the merger of two white dwarfs (ONe and/or CO)  \citep{2020A&A...644L...8O, 2021ApJ...918L..33R, 2023ApJ...944..120L}. These mergers are associated with an unusual class of supernovae known as Type Iax, which are characterized by low ejecta velocity and low luminosity \cite{2013ApJ...767...57F}. Nevertheless, as only the fifth Galactic supernova remnant for which we know the age and supernova class, Pa\,30 is worthy of study across the entire spectrum. This motivates the VERITAS observations described in these proceedings. 


\section{Observations and Analysis}
VERITAS is an array of four, 12-m aperture imaging atmospheric Cherenkov telescopes located at the Fred Lawrence Whipple Observatory in southern Arizona. Now approaching its 20th year of operations, the array is still among the most sensitive facilities for gamma-ray astronomy in the VHE region. Observations of Pa\,30 with VERITAS were conducted in \textit{wobble} mode between 2024 September 30 and 2025 January 2, with the source offset by $0.5\degree$ from the camera center. $20\U{hours}$ of good weather observations were collected, leading to a total exposure of $16.3\U{hours}$ after the removal of data with hardware issues and correction for data acquisition deadtime ($12.5\%$).

The data were analyzed and the results cross-checked using standard VERITAS analysis tools \citep{2017ICRC...35..747M, 2017ICRC...35..789C} with gamma-ray selection cuts appropriate for a source with a power-law spectral index of -2.5 (but with good sensitivity over a wide range of possible spectral indices). The reflected region method was used to estimate the residual background in the source region after selection cuts \citep{2007A&A...466.1219B}. 

\section{Results and Discussion}

\begin{figure}
  \begin{center}
    \begin{tabular}{c}
     \includegraphics[width=0.47\textwidth]{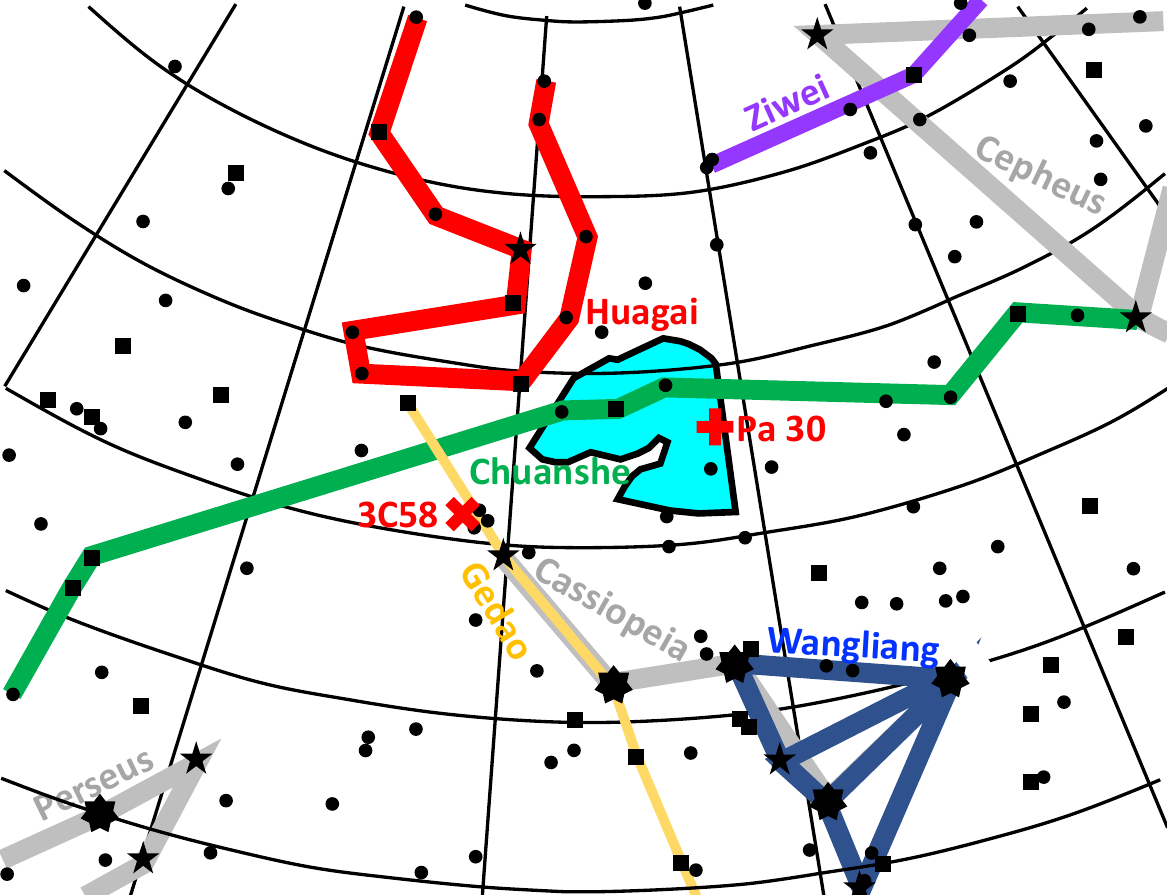}\hspace{0.5cm}\includegraphics[width=0.43\textwidth]{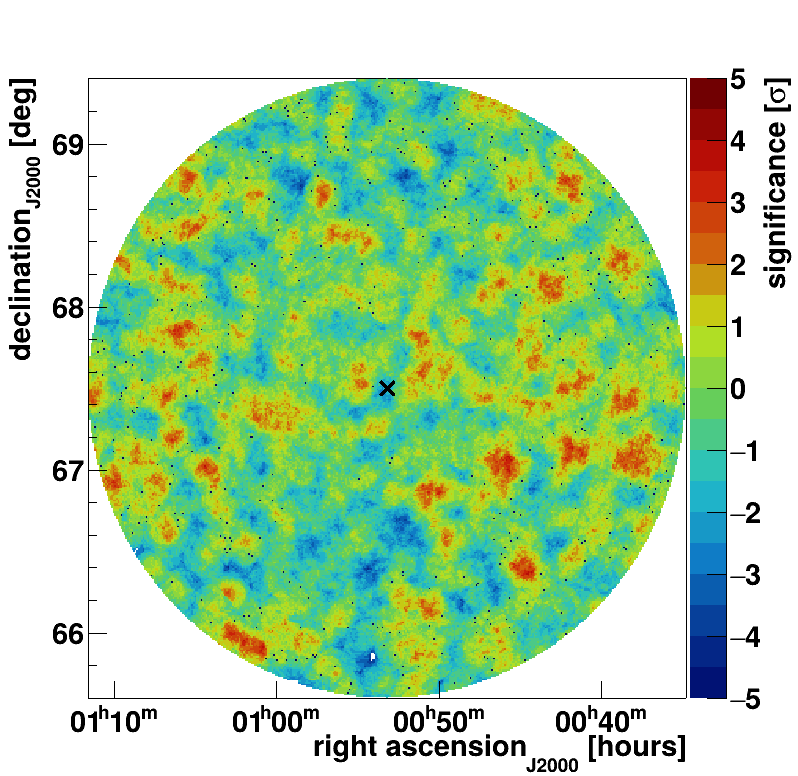} \\
    \end{tabular}
  \end{center}
  \caption  
      {
        \label{maps}
        {\bf Left:} The constraints on the position of SN\,1181 from contemporaneous Japanese and Chinese records as shown by the cyan region with black edges (figure from \citep{2023MNRAS.523.3885S}). Thick colored lines show ancient Chinese constellations. The locations of 3C58 and Pa\,30 are indicated.
         {\bf Right:} The VERITAS significance skymap in the region of Pa\,30. The location of the 3-arcminute diameter nebula is indicated by the black cross. }
\end{figure}

The VERITAS significance skymap is shown on the right of figure~\ref{maps}. No evidence for emission was found, and the integral upper limit on the photon flux is $6.52\times10^{-9}\UU{m}{-2}\UU{s}{-1}$ above $500\U{GeV}$ at 99\% confidence (approximately 1\% of the Crab Nebula flux at this energy).

We also note that there is no coincident or associated source in the Fermi LAT 14-Year Point Source Catalog (4FGL-DR4 \citep{2022ApJS..260...53A, 2023arXiv230712546B}). The closest catalog source is the unassociated object 4FGL\,J0057.5+6814 at an angular separation of $0.84\pm0.04\degree$ from Pa\,30.

\begin{figure}
  \begin{center}
    \begin{tabular}{c}
     \includegraphics[angle=90,width=0.8\textwidth]{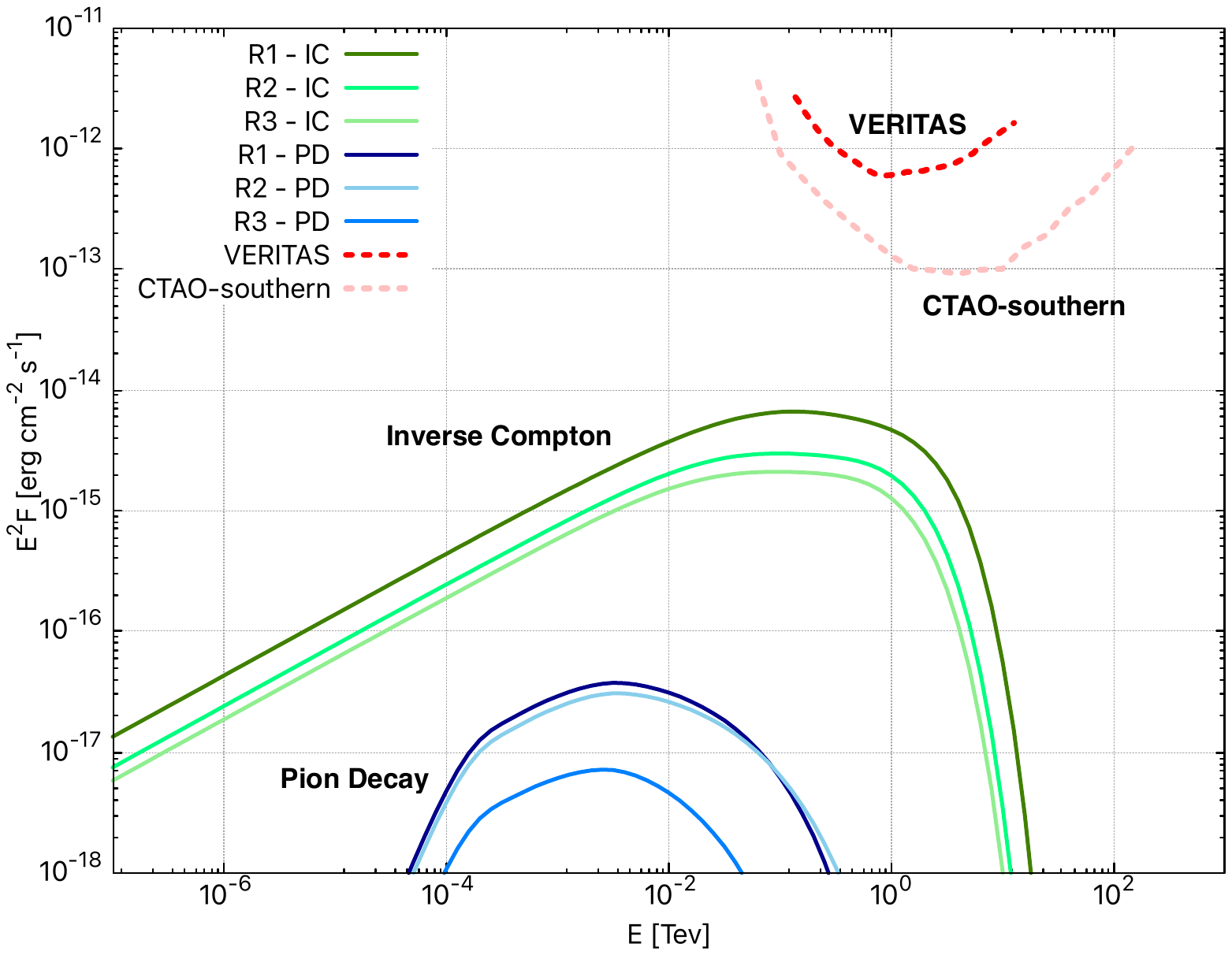}\\
    \end{tabular}
  \end{center}
  \caption  
      { Predicted gamma-ray spectra from pion decay and inverse Compton emission for the three different model configurations. Also shown are the sensitivities of VERITAS and the southern CTAO array. }
  \label{model}

\end{figure}

We have performed preliminary modeling of the system, adopting the one-dimensional, spherically symmetric framework from \citep{1998ApJ...497..807D} to describe the expansion of a Type Ia supernova remnant into a uniform ambient medium. We use the  RATPaC code \citep{2020A&A...634A..59B} to model the acceleration of cosmic rays in this scenario for three representative Type Iax supernova configurations, each varying in ejecta mass and explosion energy, as shown in table~\ref{tab:params}. The number density of the ambient interstellar medium is constrained to the low value of $0.1\UU{cm}{-3}$ by the dynamical model of Ko et al. \citep{2024ApJ...969..116K}. Figure~\ref{model} shows the predicted inverse Compton and pion-decay fluxes, suggesting that any emission in the gamma-ray band lies far below the sensitivity of VERITAS, or planned facilities such as CTAO. This can be explained as a result of the low Type Iax supernova explosion energy (hence low shock speed), coupled with the low-density ambient medium that gives few cosmic rays and little target material for hadronic interactions.

\begin{table}[h]
    \centering
    \begin{tabular}{|c|c|c|c|}
         \hline
            Model & 
            $M_{\mathrm{ejecta}}(M_{\odot}$) & $E_{\mathrm{ejecta}}$(erg) & $n_{\mathrm{ISM}}(\UU{cm}{-3})$ \\
         \hline\hline
        R1 & 0.1 & $1.5\times10^{48}$ & 0.1 \\
        \hline
        R2 & 0.6 & $2.8\times10^{48}$ & 0.1 \\
        \hline
        R3 & 1.0 & $3.375\times10^{48}$ & 0.1 \\
        \hline
    \end{tabular}
    \caption{Summary of supernova explosion parameters for the three model configurations tested.}
    \label{tab:params}
\end{table}

\acknowledgments
This research is supported by grants from the U.S. Department of Energy Office of Science, the U.S. National Science Foundation and the Smithsonian Institution, by NSERC in Canada, and by the Helmholtz Association in Germany. This research used resources provided by the Open Science Grid, which is supported by the National Science Foundation and the U.S. Department of Energy's Office of Science, and resources of the National Energy Research Scientific Computing Center (NERSC), a U.S. Department of Energy Office of Science User Facility operated under Contract No. DE-AC02-05CH11231. We acknowledge the excellent work of the technical support staff at the Fred Lawrence Whipple Observatory and at the collaborating institutions in the construction and operation of the instrument. 
\bibliographystyle{JHEP}
\bibliography{JH.bib}

\providecommand{\href}[2]{#2}\begingroup\raggedright\begin{thebibliography}{10}

\bibitem{1994A&A...287..959D}
L.O.~{Drury}, F.A.~{Aharonian} and H.J.~{Voelk}, \emph{{The gamma-ray visibility of supernova remnants. A test of cosmic ray origin}}, \href{https://doi.org/10.48550/arXiv.astro-ph/9305037}{\emph{A\&A} {\bfseries 287} (1994) 959} [\href{https://arxiv.org/abs/astro-ph/9305037}{{\ttfamily astro-ph/9305037}}].

\bibitem{2010A&A...516A..62A}
F.~{Acero}, F.~{Aharonian}, A.G.~{Akhperjanian}, G.~{Anton}, U.~{Barres de Almeida}, A.R.~{Bazer-Bachi} et~al., \emph{{First detection of VHE {\ensuremath{\gamma}}-rays from SN 1006 by HESS}}, \href{https://doi.org/10.1051/0004-6361/200913916}{\emph{A\&A} {\bfseries 516} (2010) A62} [\href{https://arxiv.org/abs/1004.2124}{{\ttfamily 1004.2124}}].

\bibitem{2011ApJ...730L..20A}
V.A.~{Acciari}, E.~{Aliu}, T.~{Arlen}, T.~{Aune}, M.~{Beilicke}, W.~{Benbow} et~al., \emph{{Discovery of TeV Gamma-ray Emission from Tycho's Supernova Remnant}}, \href{https://doi.org/10.1088/2041-8205/730/2/L20}{\emph{ApJl} {\bfseries 730} (2011) L20} [\href{https://arxiv.org/abs/1102.3871}{{\ttfamily 1102.3871}}].

\bibitem{2022A&A...662A..65H}
{H.~E.~S.~S. Collaboration}, F.~{Aharonian}, F.~{Ait Benkhali}, E.O.~{Ang{\"u}ner}, H.~{Ashkar}, M.~{Backes} et~al., \emph{{Evidence for {\ensuremath{\gamma}}-ray emission from the remnant of Kepler's supernova based on deep H.E.S.S. observations}}, \href{https://doi.org/10.1051/0004-6361/202243096}{\emph{A\&A} {\bfseries 662} (2022) A65} [\href{https://arxiv.org/abs/2201.05839}{{\ttfamily 2201.05839}}].

\bibitem{1989ApJ...342..379W}
T.C.~{Weekes}, M.F.~{Cawley}, D.J.~{Fegan}, K.G.~{Gibbs}, A.M.~{Hillas}, P.W.~{Kowk} et~al., \emph{{Observation of TeV Gamma Rays from the Crab Nebula Using the Atmospheric Cerenkov Imaging Technique}}, \href{https://doi.org/10.1086/167599}{\emph{ApJ} {\bfseries 342} (1989) 379}.

\bibitem{2001A&A...370..112A}
F.~{Aharonian}, A.~{Akhperjanian}, J.~{Barrio}, K.~{Bernl{\"o}hr}, H.~{B{\"o}rst}, H.~{Bojahr} et~al., \emph{{Evidence for TeV gamma ray emission from Cassiopeia A}}, \href{https://doi.org/10.1051/0004-6361:20010243}{\emph{A\&A} {\bfseries 370} (2001) 112} [\href{https://arxiv.org/abs/astro-ph/0102391}{{\ttfamily astro-ph/0102391}}].

\bibitem{2002ISAA....5.....S}
F.R.~{Stephenson} and D.A.~{Green}, \emph{{Historical supernovae and their remnants}}, {\emph{International Series in Astronomy and Astrophysics} {\bfseries 5} (2002) }.

\bibitem{2014A&A...567L...8A}
J.~{Aleksi{\'c}}, S.~{Ansoldi}, L.A.~{Antonelli}, P.~{Antoranz}, A.~{Babic}, P.~{Bangale} et~al., \emph{{Discovery of TeV {\ensuremath{\gamma}}-ray emission from the pulsar wind nebula 3C 58 by MAGIC}}, \href{https://doi.org/10.1051/0004-6361/201424261}{\emph{A\&A} {\bfseries 567} (2014) L8} [\href{https://arxiv.org/abs/1405.6074}{{\ttfamily 1405.6074}}].

\bibitem{2008ApJS..174..379F}
R.~{Fesen}, G.~{Rudie}, A.~{Hurford} and A.~{Soto}, \emph{{Optical Imaging and Spectroscopy of the Galactic Supernova Remnant 3C 58 (G130.7+3.1)}}, \href{https://doi.org/10.1086/522781}{\emph{ApJs} {\bfseries 174} (2008) 379}.

\bibitem{2021ApJ...918L..33R}
A.~{Ritter}, Q.A.~{Parker}, F.~{Lykou}, A.A.~{Zijlstra}, M.A.~{Guerrero} and P.~{Le D{\^u}}, \emph{{The Remnant and Origin of the Historical Supernova 1181 AD}}, \href{https://doi.org/10.3847/2041-8213/ac2253}{\emph{ApJL} {\bfseries 918} (2021) L33} [\href{https://arxiv.org/abs/2105.12384}{{\ttfamily 2105.12384}}].

\bibitem{2014apn6.confE..48K}
M.~{Kronberger}, G.H.~{Jacoby}, A.~{Acker}, F.~{Alves}, D.J.~{Frew}, D.~{Goldman} et~al., \emph{{New Planetary Nebulae and Candidates from Multicolour Multiwavelength Surveys}},  in \emph{Asymmetrical Planetary Nebulae VI Conference}, C.~{Morisset}, G.~{Delgado-Inglada} and S.~{Torres-Peimbert}, eds., p.~48, Apr., 2014.

\bibitem{2023ApJ...945L...4F}
R.A.~{Fesen}, B.E.~{Schaefer} and D.~{Patchick}, \emph{{Discovery of an Exceptional Optical Nebulosity in the Suspected Galactic SN Iax Remnant Pa 30 Linked to the Historical Guest Star of 1181 CE}}, \href{https://doi.org/10.3847/2041-8213/acbb67}{\emph{ApJl} {\bfseries 945} (2023) L4} [\href{https://arxiv.org/abs/2301.04809}{{\ttfamily 2301.04809}}].

\bibitem{2023MNRAS.523.3885S}
B.E.~{Schaefer}, \emph{{The path from the Chinese and Japanese observations of supernova 1181 AD, to a Type Iax supernova, to the merger of CO and ONe white dwarfs}}, \href{https://doi.org/10.1093/mnras/stad717}{\emph{MNRAS} {\bfseries 523} (2023) 3885} [\href{https://arxiv.org/abs/2301.04807}{{\ttfamily 2301.04807}}].

\bibitem{2020A&A...644L...8O}
L.M.~{Oskinova}, V.V.~{Gvaramadze}, G.~{Gr{\"a}fener}, N.~{Langer} and H.~{Todt}, \emph{{X-rays observations of a super-Chandrasekhar object reveal an ONe and a CO white dwarf merger product embedded in a putative SN Iax remnant}}, \href{https://doi.org/10.1051/0004-6361/202039232}{\emph{A\&A} {\bfseries 644} (2020) L8} [\href{https://arxiv.org/abs/2008.10612}{{\ttfamily 2008.10612}}].

\bibitem{2023ApJ...944..120L}
F.~{Lykou}, Q.A.~{Parker}, A.~{Ritter}, A.A.~{Zijlstra}, D.J.~{Hillier}, M.A.~{Guerrero} et~al., \emph{{A New Study on a Type Iax Stellar Remnant and its Probable Association with SN 1181}}, \href{https://doi.org/10.3847/1538-4357/acb138}{\emph{ApJ} {\bfseries 944} (2023) 120} [\href{https://arxiv.org/abs/2208.03946}{{\ttfamily 2208.03946}}].

\bibitem{2013ApJ...767...57F}
R.J.~{Foley}, P.J.~{Challis}, R.~{Chornock}, M.~{Ganeshalingam}, W.~{Li}, G.H.~{Marion} et~al., \emph{{Type Iax Supernovae: A New Class of Stellar Explosion}}, \href{https://doi.org/10.1088/0004-637X/767/1/57}{\emph{ApJ} {\bfseries 767} (2013) 57} [\href{https://arxiv.org/abs/1212.2209}{{\ttfamily 1212.2209}}].

\bibitem{2017ICRC...35..747M}
G.~{Maier} and J.~{Holder}, \emph{{Eventdisplay: An Analysis and Reconstruction Package for Ground-based Gamma-ray Astronomy}},  in \emph{35th International Cosmic Ray Conference (ICRC2017)}, vol.~301 of \emph{International Cosmic Ray Conference}, p.~747, July, 2017, \href{https://doi.org/10.22323/1.301.0747}{DOI} [\href{https://arxiv.org/abs/1708.04048}{{\ttfamily 1708.04048}}].

\bibitem{2017ICRC...35..789C}
J.~{Christiansen} and {VERITAS Collaboration}, \emph{{Characterization of a Maximum Likelihood Gamma-Ray Reconstruction Algorithm for VERITAS}},  in \emph{35th International Cosmic Ray Conference (ICRC2017)}, vol.~301 of \emph{International Cosmic Ray Conference}, p.~789, July, 2017, \href{https://doi.org/10.22323/1.301.0789}{DOI} [\href{https://arxiv.org/abs/1708.05684}{{\ttfamily 1708.05684}}].

\bibitem{2007A&A...466.1219B}
D.~{Berge}, S.~{Funk} and J.~{Hinton}, \emph{{Background modelling in very-high-energy {\ensuremath{\gamma}}-ray astronomy}}, \href{https://doi.org/10.1051/0004-6361:20066674}{\emph{A\&A} {\bfseries 466} (2007) 1219} [\href{https://arxiv.org/abs/astro-ph/0610959}{{\ttfamily astro-ph/0610959}}].

\bibitem{2022ApJS..260...53A}
S.~{Abdollahi}, F.~{Acero}, L.~{Baldini}, J.~{Ballet}, D.~{Bastieri}, R.~{Bellazzini} et~al., \emph{{Incremental Fermi Large Area Telescope Fourth Source Catalog}}, \href{https://doi.org/10.3847/1538-4365/ac6751}{\emph{ApJs} {\bfseries 260} (2022) 53} [\href{https://arxiv.org/abs/2201.11184}{{\ttfamily 2201.11184}}].

\bibitem{2023arXiv230712546B}
J.~{Ballet}, P.~{Bruel}, T.H.~{Burnett}, B.~{Lott} and {The Fermi-LAT collaboration}, \emph{{Fermi Large Area Telescope Fourth Source Catalog Data Release 4 (4FGL-DR4)}}, \href{https://doi.org/10.48550/arXiv.2307.12546}{\emph{arXiv e-prints} (2023) arXiv:2307.12546} [\href{https://arxiv.org/abs/2307.12546}{{\ttfamily 2307.12546}}].

\bibitem{1998ApJ...497..807D}
V.V.~{Dwarkadas} and R.A.~{Chevalier}, \emph{{Interaction of Type IA Supernovae with Their Surroundings}}, \href{https://doi.org/10.1086/305478}{\emph{ApJ} {\bfseries 497} (1998) 807}.

\bibitem{2020A&A...634A..59B}
R.~{Brose}, M.~{Pohl}, I.~{Sushch}, O.~{Petruk} and T.~{Kuzyo}, \emph{{Cosmic-ray acceleration and escape from post-adiabatic supernova remnants}}, \href{https://doi.org/10.1051/0004-6361/201936567}{\emph{A\&A} {\bfseries 634} (2020) A59} [\href{https://arxiv.org/abs/1909.08484}{{\ttfamily 1909.08484}}].

\bibitem{2024ApJ...969..116K}
T.~{Ko}, H.~{Suzuki}, K.~{Kashiyama}, H.~{Uchida}, T.~{Tanaka}, D.~{Tsuna} et~al., \emph{{A Dynamical Model for IRAS 00500+6713: The Remnant of a Type Iax Supernova SN 1181 Hosting a Double Degenerate Merger Product WD J005311}}, \href{https://doi.org/10.3847/1538-4357/ad4d99}{\emph{ApJ} {\bfseries 969} (2024) 116} [\href{https://arxiv.org/abs/2304.14669}{{\ttfamily 2304.14669}}].

\end{thebibliography}\endgroup





\clearpage

\begin{center} 
\section*{\centering{ VERITAS Collaboration}}

\end{center}
\noindent
A.~Archer$^{1}$,
P.~Bangale$^{2}$,
J.~T.~Bartkoske$^{3}$,
W.~Benbow$^{4}$,
Y.~Chen$^{5}$,
J.~L.~Christiansen$^{6}$,
A.~J.~Chromey$^{4}$,
A.~Duerr$^{3}$,
M.~Errando$^{7}$,
M.~Escobar~Godoy$^{8}$,
J.~Escudero Pedrosa$^{4}$,
Q.~Feng$^{3}$,
S.~Filbert$^{3}$,
L.~Fortson$^{9}$,
A.~Furniss$^{8}$,
W.~Hanlon$^{4}$,
O.~Hervet$^{8}$,
C.~E.~Hinrichs$^{4,10}$,
J.~Holder$^{11}$,
T.~B.~Humensky$^{12,13}$,
M.~Iskakova$^{7}$,
W.~Jin$^{5}$,
M.~N.~Johnson$^{8}$,
E.~Joshi$^{14}$,
M.~Kertzman$^{1}$,
M.~Kherlakian$^{15}$,
D.~Kieda$^{3}$,
T.~K.~Kleiner$^{14}$,
N.~Korzoun$^{11}$,
S.~Kumar$^{12}$,
M.~J.~Lang$^{16}$,
M.~Lundy$^{17}$,
G.~Maier$^{14}$,
C.~E~McGrath$^{18}$,
P.~Moriarty$^{16}$,
R.~Mukherjee$^{19}$,
W.~Ning$^{5}$,
R.~A.~Ong$^{5}$,
A.~Pandey$^{3}$,
M.~Pohl$^{20,14}$,
E.~Pueschel$^{15}$,
J.~Quinn$^{18}$,
P.~L.~Rabinowitz$^{7}$,
K.~Ragan$^{17}$,
P.~T.~Reynolds$^{21}$,
D.~Ribeiro$^{9}$,
E.~Roache$^{4}$,
I.~Sadeh$^{14}$,
L.~Saha$^{4}$,
H.~Salzmann$^{8}$,
M.~Santander$^{22}$,
G.~H.~Sembroski$^{23}$,
B.~Shen$^{12}$,
M.~Splettstoesser$^{8}$,
A.~K.~Talluri$^{9}$,
S.~Tandon$^{19}$,
J.~V.~Tucci$^{24}$,
J.~Valverde$^{25,13}$,
V.~V.~Vassiliev$^{5}$,
D.~A.~Williams$^{8}$,
S.~L.~Wong$^{17}$,
T.~Yoshikoshi$^{26}$\\
\noindent
$^{1}${Department of Physics and Astronomy, DePauw University, Greencastle, IN 46135-0037, USA}

\noindent
$^{2}${Department of Physics, Temple University, Philadelphia, PA 19122, USA}

\noindent
$^{3}${Department of Physics and Astronomy, University of Utah, Salt Lake City, UT 84112, USA}

\noindent
$^{4}${Center for Astrophysics $|$ Harvard \& Smithsonian, Cambridge, MA 02138, USA}

\noindent
$^{5}${Department of Physics and Astronomy, University of California, Los Angeles, CA 90095, USA}

\noindent
$^{6}${Physics Department, California Polytechnic State University, San Luis Obispo, CA 94307, USA}

\noindent
$^{7}${Department of Physics, Washington University, St. Louis, MO 63130, USA}

\noindent
$^{8}${Santa Cruz Institute for Particle Physics and Department of Physics, University of California, Santa Cruz, CA 95064, USA}

\noindent
$^{9}${School of Physics and Astronomy, University of Minnesota, Minneapolis, MN 55455, USA}

\noindent
$^{10}${Department of Physics and Astronomy, Dartmouth College, 6127 Wilder Laboratory, Hanover, NH 03755 USA}

\noindent
$^{11}${Department of Physics and Astronomy and the Bartol Research Institute, University of Delaware, Newark, DE 19716, USA}

\noindent
$^{12}${Department of Physics, University of Maryland, College Park, MD, USA }

\noindent
$^{13}${NASA GSFC, Greenbelt, MD 20771, USA}

\noindent
$^{14}${DESY, Platanenallee 6, 15738 Zeuthen, Germany}

\noindent
$^{15}${Fakult\"at f\"ur Physik \& Astronomie, Ruhr-Universit\"at Bochum, D-44780 Bochum, Germany}

\noindent
$^{16}${School of Natural Sciences, University of Galway, University Road, Galway, H91 TK33, Ireland}

\noindent
$^{17}${Physics Department, McGill University, Montreal, QC H3A 2T8, Canada}

\noindent
$^{18}${School of Physics, University College Dublin, Belfield, Dublin 4, Ireland}

\noindent
$^{19}${Department of Physics and Astronomy, Barnard College, Columbia University, NY 10027, USA}

\noindent
$^{20}${Institute of Physics and Astronomy, University of Potsdam, 14476 Potsdam-Golm, Germany}

\noindent
$^{21}${Department of Physical Sciences, Munster Technological University, Bishopstown, Cork, T12 P928, Ireland}

\noindent
$^{22}${Department of Physics and Astronomy, University of Alabama, Tuscaloosa, AL 35487, USA}

\noindent
$^{23}${Department of Physics and Astronomy, Purdue University, West Lafayette, IN 47907, USA}

\noindent
$^{24}${Department of Physics, Indiana University Indianapolis, Indianapolis, Indiana 46202, USA}

\noindent
$^{25}${Department of Physics, University of Maryland, Baltimore County, Baltimore MD 21250, USA}

\noindent
$^{26}${Institute for Cosmic Ray Research, University of Tokyo, 5-1-5, Kashiwa-no-ha, Kashiwa, Chiba 277-8582, Japan}

\end{document}